\documentstyle[12pt,aasms4]{article}

\def\1259{PSR~B1259--63}
\def\4.8{$4.8~{\rm GHz}$}
\def\8.4{$8.4~{\rm GHz}$}
\def\nud{$\Delta \nu_{\rm d}$}
\def\td{$\Delta t_{\rm d}$}
\def\stromgren{Str\"{o}mgren}
\def\HI{H{\sc i}}
\def\HII{H{\sc ii}}
\def\deg{\ifmmode^{\circ}\else$^{\circ}$\fi} 
\def\min{\ifmmode^{\prime}\;\else$^{\prime}\;$\fi}

\lefthead{McClure-Griffiths et al.}
\righthead{Scintillation of PSR B1259-63}

\begin{document}

\title{Diffractive Scintillation of the Pulsar PSR~B1259-63} 

\author{N. M. McClure-Griffiths\altaffilmark{1,}\altaffilmark{2}, Simon
Johnston\altaffilmark{3}, Daniel R. Stinebring\altaffilmark{1}, and Luciano
Nicastro\altaffilmark{4}} 

\altaffiltext{1}{Department of Physics, Oberlin College, Oberlin, OH 44074}

\altaffiltext{2}{Present address: Department of Astronomy, University of
Minnesota, Minneapolis, MN 55455}

\altaffiltext{3}{Research Centre for Theoretical Astrophysics, University of
Sydney, NSW 2006, Australia}

\altaffiltext{4}{Instituto TeSRE -- CNR, Via Gobetti 101, 40129 Bologna,
Italy}

\vspace{0.25 in}


\vspace{0.25 in}

\authoraddr{Address correspondence regarding this manuscript to: 
		Dan Stinebring
		Dept. of Physics
		Oberlin College
		Oberlin, OH 44074}

\begin{abstract}

\1259 is in a highly eccentric 3.4 yr orbit around the Be-star SS2883. The
system is located in the direction of the Sagittarius-Carina spiral arm at a
distance of $\sim$1.5 kpc.  We have made scintillation observations of the
pulsar far from periastron at \4.8 and \8.4, determining the diffractive
bandwidth, \nud, and timescale, \td, at both frequencies.  We find no
dependency on orbital phase until within 30 days of periastron. The data
indicate that the scintillation is caused, not by the circumstellar
environment, but by an \HII\ region within the Sagittarius-Carina spiral arm
located at least three-quarters of the way to the pulsar.  Close to
periastron, when the line-of-sight to the pulsar intersects the disk of the
Be star, the electron densities within the disk are sufficient to overcome
the `lever-arm effect' and produce a reduction in the scintillation
bandwidth by six orders of magnitude.

\end{abstract}

\section{Introduction}
\label{sec:intro}
Scintillation studies provide us with a powerful tool for determining
conditions along the line-of-sight to a particular pulsar. They have the
potential to determine both the strength of the scattering process and the
location of the dominant scattering screen and have been extremely
successful in determining some properties of the interstellar medium (Cordes
et al. 1991; Gupta 1995). In binary pulsars which are interacting with their
companion stars, such as the ablating millisecond pulsars, scatter
broadening provides us with a vital clue in understanding the eclipse
mechanisms (Luo \& Melrose 1995).  We present observations and
interpretation of the scintillation parameters for the binary pulsar \1259,
the only known radio pulsar with a Be star companion.

\1259 was identified as a pulsar in a highly eccentric ($e \sim 0.87$), 3.4
yr orbit with the $10~{\rm M_{\odot}}$ Be-star, SS2883 (Johnston et al.\
1992).  Optical observations determined that SS2883 is of spectral type B2e,
and also suggested that the system is at a distance $D\sim$1.5 kpc, placing
the pulsar in or just beyond the Sagittarius-Carina spiral arm (Johnston et
al.\ 1994). Be stars are characterized by a circumstellar disk of cool,
dense material, which extends outward from the equator to a few tens of
stellar radii, and a hot, low-density polar wind.

Observations of the pulsar have shown that dispersion measure (DM)
variations around the orbit are $\lesssim 0.2$ cm$^{-3}$pc, except close to
periastron where the DM increases by $\sim$15 cm$^{-3}$pc.  At this time the
pulsar undergoes extreme scatter broadening at frequencies up to at least
$5~{\rm GHz}$ (Johnston et al. 1996).  The increased DM and the scatter
broadening can be explained by the line-of-sight to the pulsar traversing
the Be star's wind and disk (Melatos, Johnston, \& Melrose 1995).  \1259
thus potentially provides us with a probe of the wind structure of a Be star
(far from periastron) and the disk structure (close to periastron).

A priori, one can think of four potential locations for the scattering
material: the general interstellar medium, dense material within the
spiral arm, an \HII\ bubble around the hot star, or the wind/disc of
the Be star.  In this paper we examine each of these possibilities in
turn.
\section{Observations and Results}
\label{sec:observ}
We made a total of 46 independent observations of \1259 between 1993 August
and 1997 February using the Parkes 64-m radio telescope. This excludes data
taken in 1993 December and 1994 January during the periastron passage of the
pulsar (Johnston et al. 1996).  The observations were made at \4.8 and \8.4
using dual-channel, cryogenically cooled systems sensitive to two orthogonal
linear polarizations.  The signals from the receivers were down-converted to
an intermediate frequency and then passed into a back-end filterbank which
consisted of a number of contiguous frequency channels.  Prior to 1996
October the filterbank consisted of 64 channels, each $5~{\rm MHz}$ wide for
a total bandwidth of $320~{\rm MHz}$.  Since 1996 October we have used an
alternative back-end system consisting of 128 channels, each $3~{\rm MHz}$
wide (total bandwidth $384~{\rm MHz}$).  The signal was detected, high and
low-pass filtered, and sampled at $0.6~{\rm ms}$.  The observations were
typically between 30 min and 100 min in duration.

The off-line data analysis was performed in two stages.  First, a dynamic
spectrum was constructed by forming pulse profiles for each frequency
channel over a small length of time, typically $10~{\rm s}$.  In order to
maximize the signal-to-noise ratio, we often summed consecutive frequency
channels and time-intervals, making sure that we retained adequate
resolution across individual scintillation maxima.  Subsequently, a
2-dimensional autocorrelation was formed of the summed dynamic spectrum, and
1-dimensional Gaussians fitted to the zero lag in time and frequency to find
the diffractive bandwidth, \nud, and diffractive time scale, \td,
respectively.  No significant tilt of the autocorrelation function was
observed, and we computed fractional uncertainties of the values based on
the number of scintles present in the dynamic spectrum following Cordes
(1986).  The scintillation parameters for each observation and their
associated errors are plotted as a function of days from periastron in
Figure~\ref{fig:scint}.

The main finding from these observations is that there is little or no
change in the scintillation parameters with orbital phase from 25 days after
periastron until 100 days before periastron. The apparent decrease in the
scintillation bandwidth at \4.8 in late 1996 and early 1997 is most likely
due to the improved spectral resolution.  The standard deviation of the
values throughout the orbit is $\sim$30\%, consistent with the individual
uncertainties at both \4.8 and \8.4.  Averaging all data at \4.8, we find
that \nud\ = 8.5 $\pm 0.9~{\rm MHz}$ and \td\ = 190 $\pm 10~{\rm s}$.  At
\8.4, we find that \nud\ = 70 $\pm 20~{\rm MHz}$ and \td\ = 360 $\pm
100~{\rm s}$.  The contrast of these values far from periastron with the
$\sim$20 Hz scintillation bandwidth at \4.8 just prior to periastron is
startling.
\section{Discussion}
\label{sec:results}
Scattering in the interstellar medium (ISM) is due to electron density
inhomogeneities which are thought to be distributed as a power-law, $P_{3N}
= C_{N}^{2} q^{-\beta}$. $P_{3N}$ is the 3-dimensional electron density
power spectrum, $C_{N}^{2}$ gives the averaged level of turbulence along a
particular line-of-sight, $q$ is the spatial wavenumber, and $\beta$ is the
spectral index of the power spectrum (e.g. Kaspi \& Stinebring 1992).  In
the discussion below we assume Kolmogorov turbulence such that $\beta =
11/3$.

Using the definition for $C_{N}^{2}$ given in Cordes (1986) and the
parameters for \1259, we find $C_{N}^{2} = 10^{-1.3}~{\rm m^{-20/3}}$.  This
value places \1259 in the top five pulsars ranked by $C_{N}^{2}$ and is more
than two orders of magnitude greater than the `canonical' value of
$10^{-3.5}~{\rm m^{-20/3}}$ for the warm, ionized (ISM).  Further evidence
for the enhanced density towards the pulsar comes from the pulsar's DM. At a
distance of 1.5 kpc, the Taylor \& Cordes (1993) model yields an expected DM
of $\sim 50~{\rm pc~cm^{-3}}$, in contrast to the measured value of $147~{\rm
pc~cm^{-3}}$.

The scintillation velocity, $V_{iss}$, can be computed from
\begin{equation}
\label{eq:viss}
V_{iss}= 3.85 \times 10^4 \frac {\sqrt{\Delta \nu _{{\rm d,MHz}} x D_{\rm
kpc} }}{\nu_{\rm GHz} \Delta t_d}~{\rm km~s^{-1}}.
\end{equation}
where $\nu$ is the observing frequency and $x$ is the ratio between
the screen-observer distance and the screen-pulsar distance (Gupta et
al.\ 1994; Gupta 1995).  Assuming a mid-placed ($x=1$) scattering
screen we derive a scintillation velocity of $\sim$150 km s$^{-1}$.
We note that the scintillation velocity, as a function of orbital phase, is constant within the errors.  The small and slowly changing
component of the pulsar's orbital velocity cannot be detected in the
data.  The proper motion of \1259 has not been measured, although
Johnston et al. (1994) estimate a recessional velocity of $80~{\rm
km~s^{-1}}$ based on the shift of the H$\alpha$ line.  Theoretical
considerations of the evolution of systems like \1259 give an upper
limit of $90~{\rm km~s^{-1}}$ for the space velocity (Brandt \&
Podsiadlowski 1995).  To reconcile this with the measured
scintillation velocity implies $x\gtrsim 3$, placing the screen closer
to the pulsar than the observer.

Evidently, then, there is good evidence for enhanced-density scattering
material located significantly closer to the pulsar than mid-way from the
observer. In the following discussion we derive an expression linking the
scintillation bandwidth with the total electron density variance along the
line-of-sight, $\delta n_e^2$, the outer scale of density inhomogeneities,
$l_o$, and the thickness of the scattering screen, $\Delta L$.  We then
consider three possible locations of the scattering screen: the wind of the
Be star, a possible \stromgren\ sphere surrounding the system, and the
Sagittarius-Carina spiral arm, all three of which satisfy the condition on
$x$.

>From geometric ray optics, we can compute the time delay, $\tau$, of
scattered rays and using the `uncertainty relationship', $2\pi \tau
\Delta\nu_{\rm d} = 1$, we have
\begin{equation}
\Delta \nu_{\rm d} = \frac{c}{\pi}\,\,\,\, \frac{(1+x)^2}{x} \,\,\,\, \frac {1}{D \theta^2 _{\rm scatt}},
\end{equation}
where $c$ is the speed of light and $\theta_{\rm scatt}$ is the
scattering angle at the screen.  (Note that for a mid-placed screen,
$x=1$, the angular broadening, $\theta$, is $\theta_{\rm scatt}/2$ and
equation 2 becomes the familiar $\Delta \nu_{\rm d}= c/(\pi D
\theta^2)$.)  By relating the electron density inhomogeneities to the
coherence length of the electromagnetic phase, $\theta_{\rm scatt}$ can be
expressed as
\begin{equation}
\theta_{\rm scatt} = \frac{c}{2\pi\nu} \,\,\,\,\,
\left[ \frac {A_{\beta}\, \nu^2} {2\pi r_e^2 c^2}
\,\,\, \frac {1}{\Delta L \, C_N^2} \right]^{-3/5}
\end{equation}
(Cordes et al. 1985), where $r_e$ is the classical electron radius.
$C_N^2$ is given by
\begin{equation}
C_N^2 = C_{\rm SM} \,\,\,\,\, \frac{\delta n_e^2}{l_o^{2/3}}.
\end{equation}
Both $A_{\beta}$ and $C_{\rm SM}$ are dimensionless constants which depend
only on $\beta$ and have values of 2.49 and 0.18, respectively, for $\beta =
11/3$ (Cordes et al. 1985, Cordes et al. 1991).  Combining these equations
and putting the parameters in convenient astronomical units we find that
\begin{equation}
\label{eq:ne2}
\delta n_e^2 = 4.0\,{\rm cm}^{-6} \,\,\,\,\,
\Delta \nu^{-5/6}_{\rm d,MHz} \,\,\,\,\,
D^{-5/6}_{\rm kpc} \,\,\,\,\, \nu^{11/3}_{\rm GHz} \,\,\,\,\,
\frac {l_{o,{\rm pc}}^{2/3}} {\Delta L_{\rm pc}} \,\,\,\,\,
\left[ \frac {(1+x)^2}{x}\right] ^{5/6}.
\end{equation}

The wind of the Be star near apastron is a possible scattering site.
However, if this were the case, one would expect a modulation in the
scintillation parameters as a function of orbital phase.  Our observations
cover a period in which the distance of the pulsar from the Be star varies
by a factor of 3; assuming a wind profile density which varies as $r^{-2}$,
this implies a change in electron density by a factor of $\sim$10. Given
that the scintillation bandwidth scales as $\Delta n_e^{-12/5}$ (from equation
[\ref{eq:ne2}]), if $\delta n_e^{2} / n_e^{2}$ and $l_o$ remain constant we
should observe a change in \nud\ by a factor $\sim$250. Furthermore, the
wind velocity is as high as $\sim$1000 km s$^{-1}$ and this would then be
the dominant velocity in the system, causing a diffractive timescale much
shorter than observed.  Finally, it is unlikely that the electron density
within the wind is large enough to cause the observed scintillation until
very close to periastron.  Assuming the scattering screen is located only a
few hundred light seconds from the pulsar and that $\l_o \approx \Delta L$
is also of this order, then $\delta n_e^{2} \sim 10^{11}$ cm$^{-6}$.  For
even a fully modulated wind with $\delta n_e^{2} \approx n_e^{2}$ the
implied electron densities are only encountered within $\pm$30 days of
periastron.

The UV flux from hot, high-mass stars can form an \HII\ bubble (or
\stromgren\ sphere) within the ISM. The interface between these two regions
can be a source of turbulence and is a possible site for scattering. We can
put limits on the emission measure of the \stromgren\ sphere from a 12-hr
synthesis image of the pulsar made using the Australia Telescope Compact
Array. The observations were made at frequencies of 1.4 and 2.4 GHz and were
gated into 8 bins synchronous with the pulsar period.  After removal of the
pulsed component from the image, the upper limit on the brightness
temperature of any residual emission is $\sim$1 K. Using standard equations
for free-free emission from \HII\ regions and assuming a temperature of
10$^{4}$ K, the upper limit on the emission measure is 650 cm$^{-6}$pc.  The
radius of a \stromgren\ sphere for a B2 star with effective temperature of
$2\times 10^4$ K is related to the electron density within the surrounding
medium by $R_s = 13 n_e^{-2/3}~{\rm pc}$ (Prentice \& ter Haar 1969).
Hence, we see that $n_e \lesssim 20$ cm$^{-3}$ and $R_s > 1.8$ pc for the
\stromgren\ sphere to be undetectable.  Applying equation (\ref{eq:ne2}),
assuming $x\sim$750 and $\l_o \approx \Delta L \approx 1$ pc, we see that
$\delta n_e^{2}$ must be larger than $\sim 3.7\times 10^4$ cm$^{-6}$ to
yield the observed scintillation bandwidth. Such a large value is
inconsistent with the upper limit on the electron density imposed by the
lack of radio emission.  Thus, the \stromgren\ sphere cannot be the source
of the observed scattering.

The Sagittarius-Carina arm intersects the line-of-sight at a distance of
$\sim$1 kpc. The spiral arm likely contains numerous, high-density \HII\
regions which can cause both excess DM and scattering.  If we estimate $l_o
\sim 1$pc for an \HII\ region and use $x=3$, equation (\ref{eq:ne2}) yields
$\delta n_e^2 \Delta L = 600$ cm$^{-6}$pc.  From the excess DM we have that
$n_e \Delta L = 100$ cm$^{-3}$pc and assuming that the condition $\delta
n_e^2 \approx n_e^2$ holds within the \HII\ region then we find $n_e \sim
6$ cm$^{-3}$ and $\Delta L \sim 10$ pc.  The derived electron density is
consistent with $\sim$6 cm$^{-3}$ and $\sim$3 cm$^{-3}$ obtained by
Heiles, Reach, \& Koo (1996) and McKee \& Williams (1997), respectively, for
`typical' \HII\ regions.  We thus conclude that the source of the enhanced
scattering is one (or several) \HII\ regions within the Sagittarius-Carina
spiral arm.

A number of other factors point to the Sagittarius-Carina spiral arm as a
dominant source of scattering. PSR~J1243--6423, only 2.2\deg\ distant from
\1259 on the sky, has very similar scintillation parameters to those of
\1259 (Johnston et al. 1997).  Its DM-derived distance is $12~{\rm kpc}$,
yet indications from \HI\ absorption suggest a distance as low as $2.2~{\rm
kpc}$ (Frail \& Weisberg 1990).  It seems evident that the spiral arm
contributes both to its large DM and relatively small scintillation
bandwidth.  Nicastro \& Johnston (1997) recently discovered a pulsar only
7\min\ from \1259 with an exceedingly high DM of $875~{\rm
pc~cm^{-3}}$. Such a large DM is not expected in this part of the Galactic
plane, and this again points to the spiral arm as a significant source of
excess electrons.

Near periastron we conclude that the disk of the Be star must dominate the
scattering. Twenty days from periastron \nud\ is $ 30~{\rm Hz}$ at $1.4~{\rm
GHz}$.  We assume that the proximity of the scattering screen implies $x
\sim 5\times 10^8$. We take $\Delta L \approx l_o \sim 100$ light seconds
and hence obtain $\delta n_e^2$ of $\sim 10^{14}$ cm$^{-6}$. Again,
assuming $\delta n_e^2 \approx n_e^2$, an electron density of $\sim 10^{7}$
cm$^{-3}$ is required.  Modeling of the disk of the Be star (Melatos et
al. 1995) shows that the maximum density encountered along the line of sight
through the disk reaches $10^{7}$ cm$^{-3}$ some 25 days prior to periastron
and rises by an order of magnitude at periastron itself. Hence, near to
periastron, there is sufficient density in the Be star's disk to produce the
observed scattering.  Because the scattering angle is $\sim$0.5\deg, the
angular broadening produces an image size of $8\times 10^{10}$ cm on the
screen. This value is significantly larger than the diffractive coherence
scale of $\sim$10$^{9}$ cm for scattering in the spiral arm. We thus expect
a quenching of the scintillation seen through the spiral arm, which is
confirmed by the observations.
\section{Conclusions}
\label{sec:conclusion}
We have determined the scintillation parameters of the binary pulsar \1259
as a function of orbital phase and conclude that there is no change in these
parameters except during a small portion of the orbit around the time of
periastron. The scintillation bandwidth is much less than expected if the
scintillation were solely due to the background ISM.  We have demonstrated
that the electron densities are not sufficiently large in either the wind of
the Be star far from periastron or a putative \stromgren\ sphere to produce
the observed scattering.  The Sagittarius-Carina spiral arm is therefore the
most likely origin of the scintillation far from periastron.  At periastron,
however, the electron density in the Be star disk is sufficiently large to
counterbalance the `lever-arm effect' of having the screen so close to the
pulsar. This produces extremely strong scattering, and the resultant
broadening of the pulsar image is sufficient to quench further scintillation
in the ISM.

\acknowledgments The back-end system used in this work was designed and
built by A. Lyne at the University of Manchester. We thank R. Manchester for
help with the observations, B. Rickett for fruitful discussions and H. Fagg
and M. McColl for receiver installation and maintenance.  NMM-G and DRS
gratefully acknowledge research support from NSF grant AST-9317605 and
hospitality and support from the Australia Telescope National Facility.  The
Parkes radio telescope is part of the Australia Telescope which is funded by
the Commonwealth of Australia for operation as a National Facility managed
by CSIRO.

\vfil\eject
\newpage
\figcaption[scint.eps]{Diffractive scintillation parameters as a function of
orbital phase.  Periastron is represented as day 0 in all plots.  Figures
(a) and (c) show diffractive bandwidth with the associated uncertainties at
\4.8 and \8.4, respectively.  Figures (b) and (d) show diffractive timescale
and associated uncertainties at \4.8 and \8.4, respectively.
\label{fig:scint}}

\end{document}